\documentclass[a4paper]{article}

\usepackage{INTERSPEECH2021}
\usepackage{graphicx}
\usepackage{color}
\usepackage{stfloats}
\usepackage{booktabs}
\usepackage{multirow}
\usepackage{bbding}
\usepackage{cite}
\usepackage{amsmath}
\usepackage{mathrsfs}
\usepackage{amsfonts}
\usepackage{bm}
\usepackage{tabularx}
\usepackage{array}
\usepackage{svg}

\title{Boosting Diffusion Model for Spectrogram Up-sampling in Text-to-speech: \\An Empirical Study}

\name{Chong Zhang, Yanqing Liu, Yang Zheng, Sheng Zhao}
\address{
  Microsoft Corporation}
\email{\{v-chongzhang, yanqliu, yanzhen, szhao\}@microsoft.com}

\begin{document}
\raggedbottom

\maketitle
\begin{abstract}

Scaling text-to-speech (TTS) with autoregressive language model (LM) to large-scale datasets by quantizing waveform into discrete speech tokens is making great progress to capture the diversity and expressiveness in human speech, but the speech reconstruction quality from discrete speech token is far from satisfaction depending on the compressed speech token compression ratio. Generative diffusion models trained with score-matching loss and continuous normalized flow trained with flow-matching loss have become prominent in generation of images as well as speech. LM based TTS systems usually quantize speech into discrete tokens and generate these tokens autoregressively, and finally use a diffusion model to up sample coarse-grained speech tokens into fine-grained codec features or mel-spectrograms before reconstructing into waveforms with vocoder, which has a high latency and is not realistic for real time speech applications. In this paper, we systematically investigate varied diffusion models for up sampling stage, which is the main bottleneck for streaming synthesis of LM and diffusion-based architecture, we present the model architecture, objective and subjective metrics to show quality and efficiency improvement.

\end{abstract}

\section{Introduction}

Recent text-to-speech (TTS) studies have made great progress with discriminative end-to-end training \cite{liu2022delightfultts,liu2021delightfultts,tan2024naturalspeech}, autoregressive language models \cite{wang2023neural,zhang2023speak,xue2023foundationtts,zhang2023speechgpt,zhan2024anygpt,sigurgeirsson2023using,ramesh2021zero} and diffusion models \cite{shen2023naturalspeech,le2024voicebox,ju2024naturalspeech,kanda2024making,vyas2023audiobox,zheng2023guided,varshavsky2024semantic,li2024cm}. Language model-based TTS has an advantage over others in terms of latency and diversity sampling, demonstrating great multimodality ability in the recent GPT-4o demo \cite{cowen2024introducing}. Language model-based TTS first quantizes the continuous waveform into discrete speech tokens via VAE \cite{kim2021conditional} or RVQ \cite{zeghidour2021soundstream,borsos2023soundstorm}, and then models single-layer or multi-layer delayed discrete tokens autoregressively. Discrete speech tokens usually have long sequences, and autoregressive (AR) models often suffer from error propagation, resulting in unstable speech outputs like repeating or skipping. Shorter speech tokens require up-sampling models, like diffusion, to recover better acoustic details. In practice, a well-designed up-sampling network plays an important role in bridging the gap between coarse-grained speech tokens and fine-grained speech frames.

Generative diffusion models trained with score-matching loss and continuous normalized flow (CNF)-based models trained with flow-matching (FM) loss have become prominent in the generation of images and audio \cite{betker2023improving}. DDPM (Denoising Diffusion Probabilistic Models) \cite{ho2020denoising} in \cite{betker2023better} is trained to convert discrete speech codes into mel-spectrograms and then fine-tune the DDPM on the autoregressive latent space from the AR model outputs. \cite{san2024discrete} presents multi-band diffusion (MBD), which generates high-fidelity samples in the waveform domain from discrete representations and can be applied to a wide variety of audio domains to replace the GAN-based decoders \cite{kumar2019melgan,kong2020hifi}. Voicebox \cite{le2024voicebox} is a conditional generative model based on flow matching \cite{lipman2022flow} which additionally conditions on frame-aligned phonetic transcripts and masked audio for audio prediction. Compared to diffusion, the flow matching model has better sampling efficiency and comparable quality.

Besides the architectural differences of diffusion models, sampling from generative diffusion or normalized flow models involves the solution of stochastic or ordinary differential equations (SDE/ODE), which demand many function evaluations (NFE) to achieve accurate results. To reduce the NFE in sampling, various approaches have been proposed. For diffusion models, some methods use knowledge distillation or try to improve the sampling trajectory, albeit requiring additional training. Others harness the special structure of the neural SDE/ODE to design dedicated solvers. For CNF models, endpoint pairing relationships are optimized to straighten the marginal probability path, and customized ODE solvers are also proposed to adaptively select the best sampling timesteps. In TTS tasks, sampling matters more due to the low tolerance of model latency. \cite{guan2024reflow} introduces a speech synthesis model based on the rectified flow model \cite{liu2022flow}, an ODE model that transports Gaussian distribution to the ground-truth mel-spectrogram distribution by straight line paths as much as possible, which surpasses the performance of most diffusion-based TTS models with just one sampling step during the inference stage and eliminates the need for pre-training a teacher model.

In this study, we utilized a pretrained TTS model (TorToise) as a framework to train a family of diffusion and flow-matching based TTS models and conducted a comparative analysis of several sampling acceleration methods, including DDIM, consistency distillation for diffusion models, and rectified flow, multisample flow, and bespoke solvers for flow-based models. We aim to provide insights into diffusion and flow matching models from a unified view, facilitating the selection of the most suitable sampling acceleration method for spectrum up-sampling in language model and diffusion-based TTS. Audio samples are available at  https://cognitivespeech.github.io/diffusionstudy.

\section{Method}

\subsection{Backbone TTS model}

We adopt TorToise TTS and its corresponding training framework, DLAS, as the backbone TTS model. TorToise consists of an autoregressive transformer and a diffusion model. The autoregressive model takes text tokens as input and outputs mel-spectrogram tokens, which the diffusion model uses as conditioning information to transform Gaussian noise back into mel-spectrograms. The mel-spectrogram is then decoded into a waveform by a pretrained vocoder. This study focuses on the diffusion model's impact on up-sampling, and the autoregressive transformer is first pretrained on a single-speaker dataset and is frozen during subsequent training.

The TorToise trick is also used here: the DDPM is first trained to convert discrete speech codes into mel-spectrograms. The DDPM or flow matching decoder is fine-tuned on the autoregressive latent space, which is pulled from the AR model outputs instead of the speech codes to have semantically rich discrete tokens, improving the efficiency of the downstream diffusion model. We also use the pretrained discriminator for speech - the contrastive language-voice pretrained transformer (CLVP) \cite{ramesh2021zero} - to rank multiple AR outputs without invoking the expensive diffusion model. The diffusion decoder of the TorToise model takes the U-Net architecture. The training strategies for each variant of the diffusion and flow-matching models are discussed below.

\subsection{Training strategy}
\subsubsection{DDPM}

The original diffusion model of TorToise is a denoising diffusion probabilistic model (DDPM) \cite{ho2020denoising}, which adopts the parameterization of noise prediction, and the loss function takes the form of Eq. (\ref{eqa:ddpm_loss}).
\begin{equation}
    \mathscr{L}(\theta) := \mathbb{E}_{t, \bm{y}, \bm{\epsilon}} \left[ \left\| \bm{\epsilon} - \bm{\epsilon_\theta} \left( \sqrt{\overline{\alpha}_t} \bm{y} + \sqrt{1 - \overline{\alpha}_t} \bm{\epsilon}, t \right) \right\|_2^2 \right]
\label{eqa:ddpm_loss}
\end{equation}
where $t\sim\mathcal{U}(0,T)$ are time steps; $\bm{y}\sim q(\bm{y})$ are  data; $\bm{\epsilon}\sim\mathcal{N}(\bm{0},\mathbf{I})$ are noises; $\bm{\epsilon}_\theta(\cdot,\cdot)$ is the diffusion decoder; and $\sqrt{\overline{\alpha_t}}$ and $\sqrt{1 - \overline{\alpha}_t}$ are variance-preserving noise schedules.

\subsubsection{EDM}

EDM \cite{karras2022elucidating} is a framework expressing diffusion models. It takes a different parameterization of time steps, schedule, noise distribution, etc. The loss function of EDM is:
\begin{equation}
    \mathscr{L}(\theta) := \mathbb{E}_{\sigma, \bm{y}, \bm{\epsilon}} \left[\lambda(\sigma)\left\|\bm{D}_\theta(\bm{y}+\sigma\bm{\epsilon},\sigma)-\bm{y}\right\|_2^2\right]
\label{eqa:edm_loss}
\end{equation}
where the noise levels are distributed according to $\ln(\sigma)\sim\mathcal{N}(P_\text{mean},P_\text{std}^2)$ and are weighted by $\lambda(\sigma)$; $\bm{D}_\theta(\cdot, \cdot)$ is a denoiser function and is preconditioned in the following form:
\begin{equation}
    \bm{D}_\theta(\bm{x};\sigma)=c_\text{skip}(\sigma)\bm{x}+c_\text{out}(\sigma)\bm{F}_\theta\left(c_\text{in}(\sigma)\bm{x},c_\text{noise}(\sigma)\right)
\label{eqa:edm_denoiser}
\end{equation}
where $\bm{F}_\theta$ is the raw neural network and can use the same structure as the diffusion decoder in DDPM; $c_\text{skip}$ modulates the skip connection, $c_\text{in}$ and $c_\text{out}$ scale the input and output magnitudes, and $c_\text{noise}$ maps the noise level $\sigma$ into a conditioning input for $\bm{F}_\theta$.

\subsubsection{Consistency Distillation}

Consistency model \cite{song2023consistency} is a family of diffusion models that can map points on any trajectory of the probability flow ordinary differential equation (ODE) to the trajectory's origin (data point), thereby theoretically allowing a one-step noise-to-data inference. Consistency models can be obtained either by distilling a pretrained diffusion model or by training from scratch. We adopt the distillation method called consistency distillation (CD). It involves three models: a teacher model \( \bm{D}_\phi \), a student model \( \bm{D}_\theta \), and an exponential moving average (EMA) of the student model \( \bm{D}_{\theta^-} \). The consistency distillation loss is defined as 

\begin{equation}
     \mathscr{L}(\theta) := \mathbb{E}_{t, \bm{y}, \bm{\epsilon}} \left[\left\|\bm{D}_\theta(\bm{x}_{i+1},t_{i+1})-\bm{D}_{\theta^-}\left(\hat{\bm{x}}_i^{\phi},t_i\right)\right\|_2^2\right]
\label{eqa:cd_loss}
\end{equation}
where $\bm{x}_{i+1}=\bm{y}+t_{i+1}\bm{\epsilon}$; $\hat{\bm{x}}_i^{\phi}$ is estimated from $\bm{x}_{i+1}$ with the teacher model $\bm{D}_\phi$ and an ODE solver.

\subsubsection{Flow Matching}

Besides diffusion models, continuous normalizing flows (CNFs) represent another family of generative models. They aim to find a vector field that transforms a noise distribution into a data distribution. Flow matching \cite{lipman2022flow} is a method used to efficiently train CNFs. It is based on constructing explicit conditional probability paths between the noise distribution and the data distribution. When the conditional probability paths are chosen to be straight lines, the flow matching loss function is as follows:
\begin{equation}
     \mathscr{L}(\theta) := \mathbb{E}_{t, \bm{y}, \bm{\epsilon}} \left[\left\|\bm{v}_\theta(t\bm{y}+(1-t)\epsilon,t)-(\bm{y}-\bm{\epsilon})\right\|_2^2\right]
\label{eqa:fm_loss}
\end{equation}
where timesteps are sampled from distribution $t\sim\mathcal{U}(0,1)$. It should be noted that in CNFs, the direction of $t$ is opposite to the convention in DDPM. $\bm{v}_\theta(\cdot,\cdot)$ is a neural network that approximates the vector field. Again, the diffusion decoder in DDPM can still be used to establish $\bm{v}_\theta$.

\subsubsection{Rectified Flow}

Although the conditional probability paths most frequently used in flow matching are straight lines, it is not guaranteed that the marginal probability paths are also straight. This is because the data points and the noises used to construct the conditional probability paths are randomly paired and may not be optimal in terms of transport cost. 

Rectified Flow \cite{liu2022flow} claims that noise and data point pair$(\bm{\epsilon},\bm{\widehat{y}})$ generated by a pretrained flow matching model has a lower transport cost compared to the initially randomly sampled pairs $(\bm{\epsilon},\bm{y})$. Therefore, $(\bm{\epsilon},\bm{\widehat{y}})$ can be used to train a new flow matching model with the following loss function:
\begin{equation}
     \mathscr{L}(\theta) := \mathbb{E}_{t, \bm{\epsilon}} \left[\left\|\bm{v}_\theta(t\bm{\widehat{y}}+(1-t)\epsilon,t)-(\bm{\widehat{y}}-\bm{\epsilon})\right\|_2^2\right]
\label{eqa:rfm_loss}
\end{equation}
where $\bm{\widehat{y}}$ is generated by a pretrained velocity field $\bm{v}_\phi$ using any ODE solver with sufficient steps.

\subsubsection{Multisample Flow}

The suboptimality of the transport cost of the marginal probability path due to random noise-data sample coupling has also been acknowledged in recent research. Multisample flow matching \cite{pooladian2023multisample, onken2021ot} is proposed as a different solution from rectified flow, aiming to reduce the transport cost between a minibatch of noise samples and data samples by establishing non-trivial couplings between data and noise samples. Its loss function, as shown below, is very similar to Eq.(\ref{eqa:fm_loss}), except that it draws noise and data samples from a joint distribution other than from two independent distributions.
\begin{equation}
     \mathscr{L}(\theta) := \mathbb{E}_{t, q(\bm{y}, \bm{\epsilon})} \left[\left\|\bm{v}_\theta(t\bm{y}+(1-t)\epsilon,t)-(\bm{y}-\bm{\epsilon})\right\|_2^2\right]
\label{eqa:mfm_loss}
\end{equation}

\subsubsection{Bespoke Solver}

A bespoke solver is a flow ODE solver \cite{shaul2023bespoke} optimized to find a transformation that simplifies sampling paths of a pretrained flow model, thereby reducing the NFE steps. During sampling, a base ODE solver is used to solve the flow ODE on the transformed probability path and timesteps. Then, the endpoint on the transformed probability path is transformed back into a generated data sample. A bespoke solver is trained using the RMSE-Bespoke loss shown below, which bounds the global truncation error.
\begin{equation}
     \mathscr{L}(\theta) := \mathbb{E}_{\bm{x}_0}\sum\limits_{i=1}^{n}M_i^\theta\left\|\bm{x}(t_i)-step_x^\theta\left(t_{i-1},x\left(t_{i-1}\right);\bm{u}_t\right)\right\|_2
\label{eqa:bse_loss}
\end{equation}

\subsection{Inference}

The data sampling processes are repeated several NFEs to transform noise into target data distribution for diffusion models. DDIM, an Euler integration solver, is used for DDPM. The Euler method is also used for EDM and the flow matching model family. For the bespoke solver, the Euler ODE solver is applied to the transformed probability path.

\section{Experiment}
\subsection{Dataset}

We use an internal single-speaker dataset to fine-tune the pretrained diffusion and LM model. This dataset contains 6,000 pieces of (text, mel-spectrogram) data samples. The total audio length corresponds to about 7 hours, with a sample rate of 24000 Hz. The mel-spectrogram is extracted with 100 channels, a frame size of 1024, and a hop size of 256. The test and evaluation sets are randomly split from the training set.

\subsection{Implementation Details}

The diffusion decoder mainly consists of stacked diffusion layers, each of which comprises a ResNet convolution block and an attention block. Here, the number of diffusion layers is 10, and the number of model channels is 1024.

The models are trained for 64,000 iterations on a single NVIDIA V100 GPU with a batch size of 16. It should be noted that a batch size of 64 is used when training the multisample flow model to achieve a better coupling relationship. The Adam optimizer is adopted with an initial learning rate of 1e-4. A multi-step learning rate decay schedule is adopted with a decay rate of 0.5 at 6400, 12800, 25600, and 38400 iterations.

\subsection{Performance}

\subsubsection{MOS \& CMOS}

The evaluation set balances long sentences, short sentences, question sentences, etc. Subjective metrics include mean opinion score (MOS) and comparative mean option score (CMOS). Audio generated is sent to a human rating system where each sample is rated by at least 15 raters on a scale from 1 to 5 for MOS. For CMOS, the raters are asked to give a score ranging from -3 to 3. Different evaluation results from diffusion and flow matching models with different NFEs differ much in voice quality, hence we only pick one representative group for general subjective assessment to reduce test cost; others rely on objective tests in the following sections.

Among varied settins, flow matching balances better at latency and audio quality, as shown in Table \ref{table:mos}, the MOS of the utterances sampled at 200 steps by the flow matching model is 4.38, marginally lower than that of the ground truth, which stands at 4.46. This indicates the flow matching model’s capability to produce samples that are realistic and natural.

\begin{table}[h!]
\centering
\caption{MOS compare between flow matching and ground truth}
\begin{tabular}{>{\centering\arraybackslash}p{0.5\linewidth} >{\centering\arraybackslash}p{0.3\linewidth}}
\toprule
Method & Mean Score\\
\midrule
Ground Truth      & 4.46 \\
Flow@200   & 4.38 \\

\bottomrule
\end{tabular}

\label{table:mos}
\end{table}

The CMOS results are shown in Table \ref{table:cmos}. Note that the number after @ in the table represents the NFE used for sampling from the model. The bespoke solver model and the rectified flow model outperform the flow matching model with mean scores of 0.121 and 0.039, respectively. The multisample flow model and rectified flow model are tied, with a mean score of -0.008. In general, the three improved models of the flow matching model still achieve similar or slightly better sampling quality than the original model while greatly reducing the number of inference steps.

\begin{table}[h!]
\centering
\caption{CMOS of Flow, ReFlow and Bespoke}
\begin{tabular}{>{\centering\arraybackslash}p{0.5\linewidth} >{\centering\arraybackslash}p{0.3\linewidth}}
\toprule
Old v.s. New & Mean Score\\
\midrule
Flow@200 v.s. ReFlow@3      & 0.039 \\
ReFlow@3 v.s. MultiFlow@3   & -0.008 \\
Flow@200 v.s. Bespoke@5    & 0.121 \\

\bottomrule
\end{tabular}

\label{table:cmos}
\end{table}

\subsubsection{FSD}

We adapt the same metric for speech diversity as \cite{le2024voicebox}, referred to as Fréchet Speech Score (FSD), which is derived from Fréchet Inception Score (FID) and is widely adopted for image generation evaluations, capturing the similarity between generated and ground truth images at the distribution level, a shorter distance implies the distributions are more similar and generally reflects both higher sample quality and diversity.

It can be observed that the FSD of the original DDPM-DDIM is quite large when the NFE is very few, but it decreases rapidly as NFE increases, stabilizing around 2.2. In contrast, the FSD of EDM also decreases with increasing NFE, but it is much smaller when NFE is between 1 and 4. The consistency model further reduces the FSD at very few NFEs (1-4), but it increases with higher NFEs. This unexpected phenomenon warrants further investigation in future work. Compared to the diffusion models, the flow matching model family exhibits relatively low FSD across different NFEs and demonstrates greater stability. Among these, the multisample flow matching model performs best, and the bespoke solver model also performs well at 5 and 8 NFEs.


\begin{table*}[t]
\centering
\caption{FSD compare with different NFE}
\begin{tabular}{c*{15}{c}}
\toprule
 & & & & & & & NFE & & & & & & & \\ 
\cmidrule(lr){2-16}
         Method & 1 & 2 & 3 & 4 & 5 & 6 & 8 & 10 & 20 & 30 & 40 & 50 & 60 & 80 & 100 \\
\midrule
Flow      & 2.24& 2.27& 2.38& 2.33& 2.22& 2.23& 2.24& 2.25& 2.3& 2.35& 2.31& 2.31& 2.32& 2.33& 2.31 \\
ReFlow    & 2.13& 2.32& 2.32& 2.27& 2.3& 2.3& 2.47& 2.42& 2.25& 2.4& 2.41& 2.36& 2.34& 2.28& 2.27 \\
MultiFlow & 2.01& 2.13& 2.06& 2.08& 2.06& 2.16& 2.09& 2.14& 2.12& 2.07& 2.08& 2.07& 2.08& 2.06& 2.10 \\
Bespoke   &      &      &      &      & 2.19 &      & 2.19 &      &      &      &      &      &      &      &      \\
DDPM-DDIM & 87.76& 70.37& 2.97& 2.33& 2.34& 2.22& 2.05& 2.08& 2.05& 2.15& 2.13& 2.37& 2.36& 2.22& 2.42 \\
EDM       & 3.98& 2.72& 2.63& 2.25& 2.37& 2.37& 2.32& 2.34& 2.45& 2.41& 2.36& 2.37& 2.36& 2.41& 2.34 \\
CD        & 2.14& 2.55& 2.51& 2.29& 2.36& 2.52& 2.53& 2.71& 2.9& 2.88& 2.91& 2.99& 2.9& 2.96& 2.93 \\

\bottomrule
\end{tabular}

\label{table:fsd}
\end{table*}
\subsubsection{Similarity}

Similarity is measured by the WavLM-TDCNN speaker embedding model \cite{chen2022wavlm} between the embedding of generated speech and that of the audio context. Similar to their performance on FSD, the models' performance on similarity also highlights the differences between diffusion models and flow matching models. DDPM-DDIM and EDM exhibit an increasing trend in similarity as NFE increases. The consistency model performs well at few NFEs, but its performance decreases at higher NFEs. The flow matching models maintain steady performance overall. Among these, rectified flow, multisample flow, and the bespoke solver all improve upon the original flow matching model to varying degrees, with rectified flow performing the best.

\begin{table*}[t]
\centering
\caption{Similarity difference with different NFE}
\begin{tabular}{c*{15}{c}}
\toprule
 & & & & & & & NFE & & & & & & & \\ 
\cmidrule(lr){2-16}
         Method & 1 & 2 & 3 & 4 & 5 & 6 & 8 & 10 & 20 & 30 & 40 & 50 & 60 & 80 & 100 \\
\midrule
Flow      & 0.72& 0.76& 0.76& 0.75& 0.75& 0.75& 0.75& 0.74& 0.73& 0.73& 0.73& 0.73& 0.73& 0.72& 0.72 \\
ReFlow    & 0.73& 0.75& 0.76& 0.76& 0.76& 0.76& 0.76& 0.76& 0.76& 0.76& 0.76& 0.76& 0.76& 0.76& 0.76 \\
MultiFlow & 0.73& 0.74& 0.74& 0.74& 0.75& 0.75& 0.75& 0.74& 0.74& 0.74& 0.73& 0.73& 0.73& 0.73& 0.73 \\
Bespoke   &      &      &      &      & 0.74 &      & 0.74 &      &      &      &      &      &      &      &      \\
DDPM-DDIM & -0.07& -0.08& 0.66& 0.72& 0.73& 0.72& 0.73& 0.73& 0.72& 0.72& 0.72& 0.7& 0.7& 0.7& 0.7 \\
EDM       & 0.62& 0.7& 0.72& 0.74& 0.74& 0.73& 0.74& 0.73& 0.73& 0.73& 0.73& 0.73& 0.73& 0.73& 0.73 \\
CD       & 0.7& 0.72& 0.72& 0.72& 0.71& 0.7& 0.69& 0.68& 0.67& 0.66& 0.65& 0.65& 0.65& 0.65& 0.65 \\
\bottomrule
\end{tabular}

\label{table:sim}
\end{table*}

\subsubsection{Discussion}
The results of various evaluation indicators show that rectified flow, multisample flow, and the bespoke solver all offer different degrees of improvement compared to the original flow matching model. However, these three methods have certain practical limitations. Rectified flow requires the complete solving of the ODE using a pre-trained model in every training step to obtain the generated sample, which demands significant time and computing resources. The bespoke solver also needs to completely solve the ODE to obtain the probabilistic path target, and it must calculate and accumulate the loss for multiple points on the path, posing a greater challenge to computing resources. Additionally, it can only train for a fixed number of inference steps each time, making it less convenient to balance the NFEs and audio quality. Multisample flow incurs the least overhead during training, as it only needs to optimize the transport cost of noise and data on a mini batch. In theory, the larger the batch size, the closer the coupling relationship is to the true optimum, so in practice, it is best to use a larger batch size. Considering the overall model performance and training difficulty, Multisample flow is the best option.

Diffusion models retain their unique advantages. For instance, DDPM-DDIM exhibits the best FSD performance after NFE surpasses 8. Furthermore, despite employing the ODE sampling method in this study, the diffusion model inherently permits SDE sampling, which could potentially offer superior sampling quality in cases where the model is not perfectly trained, as discussed in \cite{nie2023blessing}.

The consistency model performs well with very few NFEs. However, during the model distillation process, we observed instability, and the loss would soar in the later stages, so we adopted an early stopping strategy. The performance degradation of the consistency model with more NFEs may be related to this, and further inspection will be conducted later.

In addition, it should be noted that this study uses a single-speaker dataset with a small amount of data and low diversity, so various models can achieve high sampling quality with fewer inference steps. If a multi-speaker dataset is used, the model may require more inference steps, and the differences between different models may increase. Moreover, the impact of guidance also needs to be taken into account. These aspects will be studied in future work.


\section{Conclusions}

In this study, we experiment with various methods, ranging from score-matching-based generative diffusion models to continuous normalized flow (CNF) models with vector field loss, to illustrate spectrum up-sampling from discrete speech tokens in LLM and diffusion-based TTS systems. Our aim is to provide guidance for future efficiency studies of diffusion-based TTS. Experimental results indicate that distillation-based methods and flow matching models can offer a better balance between speed and quality. For future work, we plan to further investigate the impact of first byte latency with diffusion or flow matching models for natural and expressive TTS.

\bibliographystyle{IEEEtran}

\bibliography{mybib}


\end{document}